\documentclass[doublecol]{epl2}
\usepackage{graphicx}
\usepackage{amssymb}
\title{Work and dissipation fluctuations near the stochastic resonance of a colloidal particle}

\author{Pierre Jop, Artyom Petrosyan \& Sergio Ciliberto}
\institute{Universit\'e de Lyon, Laboratoire de Physique de
l'\'Ecole Normale Sup\'erieure de Lyon, CNRS UMR 5276, , 46
all\'ee d'Italie, 69364 Lyon cedex 7, France.}

 \abstract{We study experimentally the
fluctuations of the injected and dissipated  energy in a system of
a colloidal particle trapped in a double well potential
periodically  modulated by an external perturbation. The work done
by the external force and the dissipated energy are measured close
to the stochastic resonance where  the injected power is maximum.
We show that the steady state fluctuation theorem holds in this
system.}

\begin{document}

\maketitle

A colloidal particle, confined in a double well potential, hops
between the two wells at a rate $r_k$, named the Kramers' rate,
which is determined by the height $\delta U$ of the energy barrier
between the two wells, specifically $r_k=\tau_o^{-1}
\exp({-{\delta U \over k_BT}})$, where $\tau_o$ is a
characteristic time,  $k_B$ the Boltzmann constant and $T$ the
heat bath temperature \cite{libchaber}.   When the double well
potential $U$ is modulated by an external periodic perturbation
whose frequency is close to $r_k$ the system presents the
stochastic resonance phenomenon\cite{Benzi}, i.e. the hops of the
particle between the two wells synchronize with the external
forcing. The stochastic resonance has been widely studied in many
different systems and it has been shown to be a bona fide
resonance looking at the resident time\cite{Benzi,gammaitoni95}
and the Fourier transform of the signal for different noise
intensity\cite{babic04}. Numerically, the stochastic resonance has
been characterized by computing the injected work done by the
external agent as a function of noise and frequency
\cite{iwai01,dan05}. However the fluctuations of the injected and
dissipated power at the stochastic resonance have never been
studied experimentally. This is a very important and general issue
within the context of Fluctuation Theorems (FT) which constitute
extremely useful relations for characterizing the probabilities of
observing entropy production or consumption in out of equilibrium
systems. These relations were first observed in the simulations of
a sheared fluid\cite{evans} and later proven both for chaotic
dynamical systems \cite{gallavotti95} and for stochastic
dynamics\cite{kurchan98}. These works lead to different
formulations which find powerful applications for measuring
free-energy difference in biology (see e.g. \cite{ritort06} for a
review). The hypothesis and the extensions of fluctuation
theorems\cite{Cohen}
 have been tested in various
 experimental systems such as colloidal particles\cite{blickle06,wang02,imparato},
 mechanical
 oscillators\cite{joubaud07}, electric circuits\cite{garnier} and optically driven
single two-level systems\cite{schuler05}.  The effect of
nonharmonic potential on the motion of a colloidal particle has
been  tested by Blickle \etal
\cite{blickle06,seifert07,schuler05}. However  the kind of
nonharmonic potential used in these experiments  did not induce a
bistable dynamics of the Brownian particle, and as far as we know
there  is only one experimental study of FT for bistable systems
which are not thermally activated  \cite{schuler05}. Recently a
numerical study, which has explored the distributions of the
dissipated heat and of the work in a Langevin dynamics  near the
stochastic resonance, has shown  that  FT holds in the long-time
limit \cite{saikia07,sahoo07}.
%The effect of a non linear
%potential on a periodically modulated system has been also
%recently theoretically discussed in ref.\cite{Dykman}

To give more insight into this  problem we  study experimentally
the Steady State Fluctuation Theorem (SSFT) in the case of a
colloidal bistable system, composed by a Brownian particle trapped
in a double well potential periodically modulated by an external
driving force. We measure the energy injected into the system by
the sinusoidal perturbation and we analyze the distributions of
work and heat fluctuations. We find that although the dynamics of
the system is strongly non-linear the SSFT holds for the work
integrated on time intervals which are only a few periods of the
driving force.

The experimental setup is composed by a custom built vertical
optical tweezers made of an oil-immersion objective (63$\times$,
N.A.=1.3) which focuses a laser beam (wavelength
$\lambda=1064$~nm) to the diffraction limit
  for trapping glass beads ($2~\mu$m in diameter). The silica beads are dispersed
 in bidistilled water in very small concentration. The suspension is introduced
 in the sample chamber of dimensions $0.25\times10\times10~mm^3$,
 then a single bead is trapped and moved away from others.
The position of the bead is tracked using a fast-camera with the
resolution of 108 nm/pixels which gives after treatment the
position of the bead with an accuracy better than 20~nm.
 The trajectories of the bead are sampled at
50~Hz.

 The position of the trap can be easily displaced on the
focal plane of the objective by deflecting the laser beam using an
acousto-optic deflector (AOD). To construct the double well
potential the laser is focused alternatively at two different
positions at a rate of $5kHz$. The residence times $\tau_i$ (with
$i=1,2$) of the laser in each of the two positions determine the
mean trapping strength felt by the trapped particle. Indeed if
$\tau_1=\tau_2=100\mu s$ the typical diffusion length of the bead
during this period is only  5~nm. As a consequence the bead feels
an  average   double-well potential: $U_0(x)=ax^4-bx^2-dx$, where
$a$, $b$ and $d$ are determined by the laser intensity and by the
distance of the two focal points. In our experiment the distance
between the two spots is  $1.45~\mu$m, which produces a trap whose
minima are at $x_{min}=\pm 0.45\mu m$. The total intensity of the
laser is $58~mW$ on the  focal plane which corresponds to an
inter-well barrier  energy $\delta U_o=1.8~k_BT$. Starting from
the static symmetric double-trap, ($\tau_1=\tau_2$) we modulate
the depth of the wells at low frequency by modulating the
residence times ($\tau_i$) during which the spot remains in each
position\footnote{We
 keep the total intensity of the laser constant
  in order to produce a more stable potential.}.
 The modulation of the average intensity is harmonic at frequency $f$ and its  amplitude
 $(\tau_2-\tau_1)/(\tau_2+\tau_1)$,
  is $0.7~\%$ of the average intensity in the static symmetric case.
 Thus the potential felt by the bead has the following profile in the axial direction:
\begin{equation}
U(x,t)=U_0(x)+U_p(x,t)=U_0+c\ x \ \sin(2 \pi f t),
\end{equation}
with $ax_{min}^4= 1.8\  k_BT$, $bx_{min}^2=3.6 \ k_BT$,
$d|x_{min}|=0.44\ k_BT$ and $c|x_{min}|=0.81~k_BT$. The amplitude
of the time dependent perturbation is synchronously acquired with
the bead trajectory.\footnote{The parameters given here are
average parameters since the coefficients $a$,
 $b$ and $c$ ,obtained from fitted steady distributions at given phases,
 vary with the phase ($\delta a/a\approx10\%$, $\delta b/b\approx\delta c/c\approx5\%$).}

\begin{figure}[htbp]
\begin{center}
\includegraphics[width=8cm]{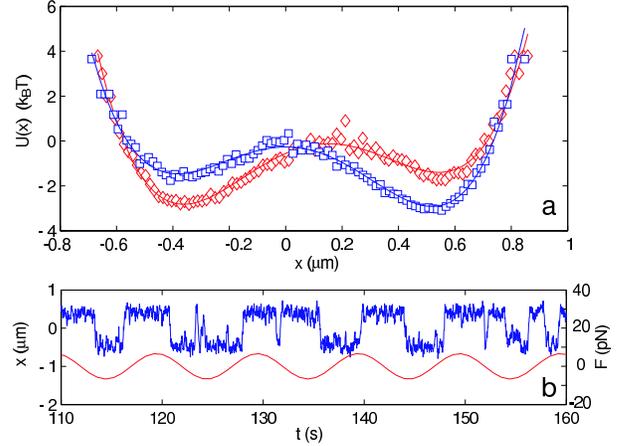}
\caption{a) The perturbed potential at $t=\frac{1}{4 f}$ and half
a forcing period later. b) Example of trajectory of the glass bead
and the corresponding perturbation at $f=0.1$ Hz.}
\label{fig:artforcingex}
\end{center}
\end{figure}

An example of the measured potential for $t=\frac{1}{4 f}$ and
$\frac{3}{4 f}$ is shown on the Fig.~\ref{fig:artforcingex}a).
This figures is obtained by measuring the probability distribution
function $P(x,t)$ of $x$ for  fixed values of $c \sin(2 \pi f t)$,
it follows that $U(x,t)=-\ln(P(x,t))$.

The $x$ position of the particle can be  described by a Langevin
equation:
\begin{equation}
    \gamma\dot x=\frac{dU}{dx}+\xi,
\end{equation}
with $\gamma=1.61 \ 10^{-8}N\ s \ m^{-1}$ the
friction coefficient and $\xi$ the stochastic force. The natural
Kramers' rate ($c=0$) for the particle is $r_k=0.3Hz$ at $T=300K$.
When $c\ne0$  the particle can experience a stochastic resonance
when the forcing frequency is close to the Kramers' rate. An
example of the sinusoidal force with the corresponding position
are shown on the figure \ref{fig:artforcingex}b).
Since the synchronization is not perfect, sometimes the particle
receives energy from the perturbation,
sometimes the bead moves against the perturbation leading to a negative work on the system.

In the following, all energies are normalized by $k_BT$. From the
trajectories, we compute  the stochastic $W_s$ and the classical
$W_{cl}$ works done by the perturbation on the system and the heat
$Q$ exchanged with the bath. These three quantities  are defined
by the following equations as  in ref.\cite{sekimoto96}:
\begin{eqnarray}
    W_s[x(t)]=\int^{t_0+t_f}_{t_0}{dt\frac{\partial U(x,t)}{\partial t}}&&  \nonumber \\
    W_{cl}[x(t)]=-\int^{t_o+t_f}_{t_0}{dt\dot x\frac{\partial U_p(x,t)}{\partial x}}&&
    \label{eq:integr}\\
    Q[x(t)]=-\int^{t_0+t_f}_{t_0}{dx\frac{\partial U(x,t)}{\partial x}}&& \nonumber
\end{eqnarray}
where in this case $t_f={n \over f}$ is a multiple of the forcing
period. We use both $W_s$ and $W_{cl}$ because they give
complementary informations on the fluctuations of the energy
injected by the external perturbation into the system (see ref.
\cite{Taniguchi} and reference therein for a discussion on this
point). For example, as we will see later, $W_s/T$ is the total
entropy production rate in this specific case \cite{seifert07b}.
The heat and the work, defined in eq.\ref{eq:integr}, are related
 through the first principle of thermodynamics: $Q=-\Delta U+W_s$,
  where $\Delta U=U(x(t_f+t_0),t_0+t_f)-U(x(t_0),t_0)$, whereas the two works
  are related by a boundary term $W=-\Delta U_p+W_s$,
  where $\Delta U_p=U_p(x(t_f+t_0),t_f+t_0)-U_p(x(t_0),t_0)$.
% Since the characteristic time evolution of the perturbation is small compared to the fluctuation of
% position and due to the harmonic form of the perturbation, the integrals are computed as follows:
%\begin{eqnarray}
%    W_s[x(t)]&=&\omega \ c \ \delta t \ \sum_{i=1}^{t_f/\delta t}x(i)\cos(\omega (t_o+t_i))  \nonumber\\
%    W_{cl}[x(t)]&=&-\Delta U_p+W_s
%    \label{eq:wqcompute}\\
%    Q[x(t)]&=&-\Delta U+W_s  \nonumber
%\end{eqnarray}
%where $\delta t$ is the sampling time.
% We checked that the direct computing of integrals of $Q$ and
%$W_{cl}$ gives the same results.
\begin{figure}[htbp]
\begin{center}
\includegraphics[width=8cm]{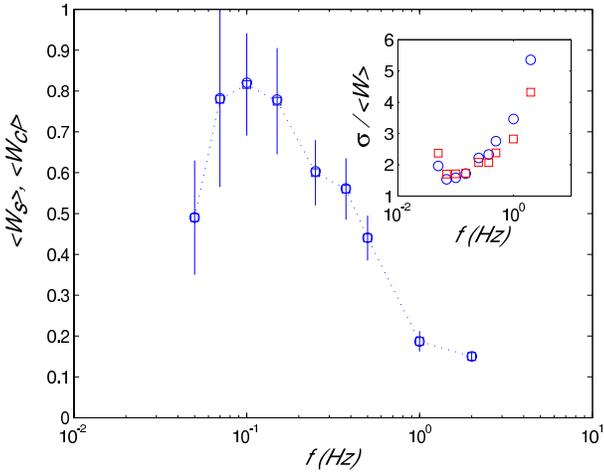}
\caption{Injected energy in the system over a single period as a
function of the driving frequency ($W_s$ $\Box$ and $W_{cl}$
$\circ$). The error bars are computed from the standard deviation
of the mean over different runs. Inset:  Standard deviations of
work distributions over a single period normalized by
 the average work as a function of the frequency (same symbols).}
\label{fig:mworkvsomega}
\end{center}
\end{figure}
We first measure the average work received over one period for
different frequencies ($t_f={1\over f}$ in eq.~\ref{eq:integr}).
Each trajectory is here recorded during 3200~s in different
consecutive runs, which corresponds to 160 up to 6400 forcing
periods, for the range of frequencies explored. In order to
increase the statistics we consider  $10^5$ different $t_o$. The
figure~\ref{fig:mworkvsomega} shows the evolution of the mean work
per period for both definitions of the work. First, the input
average work decreases to zero when the frequency tends to zero.
Indeed, the bead hops randomly several times between the two wells
during the period. Second, in the limit of high frequencies, the
particle has not the time to jump on the other side of the trap
but rather stays in the same well during the period, thus the
input energy is again decreasing when increasing frequency. In the
intermediate regime, the particle can almost synchronize with the
periodical force and follows the evolution of the potential. The
maximum of injected work is found around the frequency $f \approx
0.1$~Hz, which is comparable with half of the Kramers' rate of the
fixed potential $r_K=0.3$~Hz. This maximum of transferred energy
shows that the stochastic resonance for a Brownian particle is a
bona fide resonance, as it was previously shown in experiments
using resident time distributions \cite{gammaitoni95,schmitt06} or
directly in simulations \cite{iwai01,dan05}.
 It is worth noting that the average values of work in this case do not
 depend on their definitions: only the boundary terms, which
 vanish in average with time, are different.
 In the inset of figure~\ref{fig:mworkvsomega},
 we plot the normalized standard deviation of work distributions
 ($\sigma/\left<W\right>$) as a function of the forcing frequency.
 The curves present a minimum at the same frequency of 0.1~Hz,
 in agreement again with the resonance phenomena. However,
 we observe a difference between the two quantities.
 This underlines that these measured works have not identical fluctuations,
 which will be studied in detail in the following of the article.
\begin{figure}[htbp]
\begin{center}
\includegraphics[width=8.4cm]{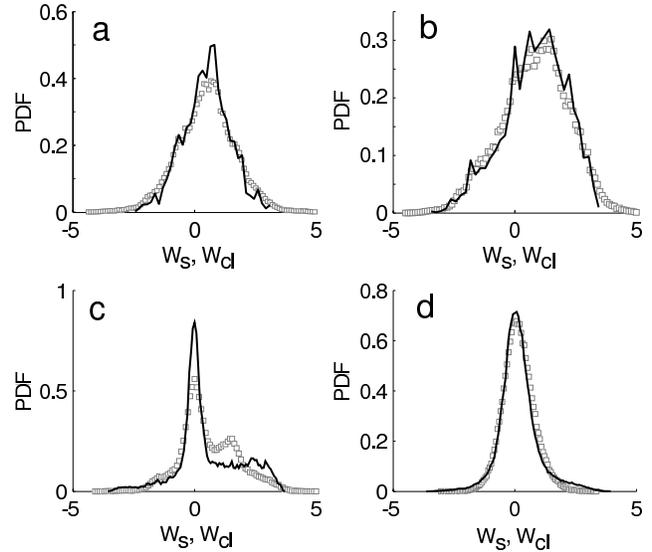}
\caption{Distributions of work over one period of $W_{cl}$
(symbols) and $W_s$
 (solid line) for a forcing at a) 0.05 Hz, b) 0.1 Hz, c) 0.375 Hz and d) 2 Hz.}
\label{fig:distrworkomeg}
\end{center}
\end{figure}
We now focus on the distributions of work over a single period.
On the figures \ref{fig:distrworkomeg}, we present
 the probability density function (PDF)  of $W_s$ and $W_{cl}$ for different frequencies.
 One can notice  the presence of a  single bump around the mean
 value for low frequency (Fig. \ref{fig:distrworkomeg}a),
 in agreement with the behavior of the bead.
  The fluctuations reach values, which are larger than six times the average injected work.
  Due to the lack of events during an experiment at such low frequency,
  we are not allowed to conclude about the shape of the distribution.
  However, it seems to tend to a gaussian distribution with deviations for large fluctuations.
Close to the resonance frequency (Fig. \ref{fig:distrworkomeg}b),
 the distribution is wide and shifted toward a larger value.
Then, when the frequency increases (Fig. \ref{fig:distrworkomeg}c),
the distribution shows several peaks.
For the classical work, the first peak, centered around zero, corresponds to
 the work received when the bead does not leave its well during the period.
 The other one corresponds to the work done on the system during a single jump of the bead.
  For higher frequencies (Fig. \ref{fig:distrworkomeg}d), only
  the central peak remains and increases,
  the distributions become more symmetrical and tends to a gaussian centered on zero since
  the bead hardly leaves its well and thus explores only the energy landscape of one well.
Although the mean works are equal, the distribution of $W_s$ and
$W_{cl}$ present significant differences close to the stochastic
resonance. In order to study in more details these distributions
of work and heat dissipation, we choose a frequency of external
driving (0.25~Hz) which  ensures a good statistic,
 by allowing the observation of the system over a sufficient number of
 periods, and  which, at the same time, produces a non trivial
 distribution
(see Fig.~\ref{fig:distrworkomeg}).
\begin{figure}[htbp]
\begin{center}
\includegraphics[width=6cm]{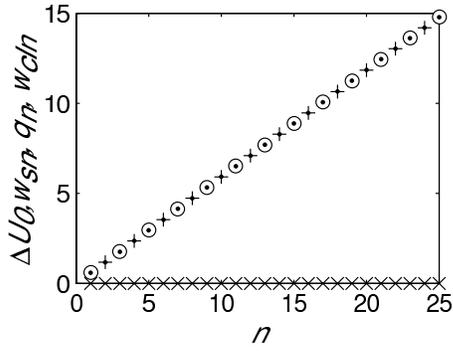}
\caption{Evolution of the mean value of works($\circ$,
\textbullet), dissipation ($+$) and potential variations
($\times$) over an increasing number of period $n$ ($f=0.25$~Hz).}
\label{fig:w2qn}
\end{center}
\end{figure}
We compute the works and the dissipation using $1.5 \  10^6$
different $t_o$ on time series which spans about 7500 period of
the driving.
%Eq.~\ref{eq:wqcompute}.
The quantities $W_{sn}$, $W_{cln}$ and $Q_n$ refer to averages
made over $n$ periods of the driving. The exchanged heat is
defined from the first principle of the dynamic: $\Delta
U_0=W_{cl}-Q$. The figure \ref{fig:w2qn} shows first the linear
increase of $W_{sn}$, $W_{cln}$ and $Q_n$ with $n$  and second
that all these quantities are equals because $\Delta U_0$ vanishes
in average (shown by the crosses in Fig.~\ref{fig:w2qn}).
\begin{figure}[htbp]
\begin{center}
\includegraphics[width=8cm]{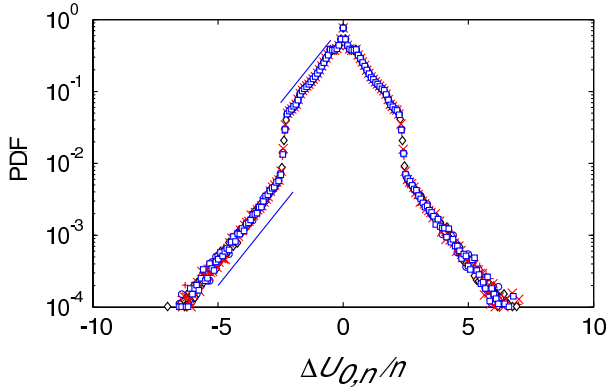}
\caption{Distribution of $\Delta U_0$ for $n=1, 2, 4, 8, 12$. The
lines are proportional to $\exp(-\Delta U_0)$ ($f=0.25$~Hz).}
\label{fig:distrdelU}
\end{center}
\end{figure}
The distribution of  $\Delta U_0$ is plotted in
Fig.~\ref{fig:distrdelU}. We observe first that these
distributions do not depend on the number of integration period.
This is due to the fact that the potential depends only on the
position at the end of the $n$-$th$ period,
 which has the same distribution for all $n$. We notice the classical large exponential wings
 preceded by a sudden decrease in the distribution. This breakdown corresponds
 to the potential energy
 of the inter-well barrier and could be understood as follow:
 The bead can not explore so often
 the high values of the potential as
 in a single well configuration but jumps likely into the other well.
 We can also remark secondary peaks around zero,
 which correspond to a small difference of energy between the two wells.
 Each peak correspond to the probability of having a single jump during driving period.
 For small values of $\Delta U_0$, the PFD decreases also exponentially,
  describing the evolution of the bead in a single trap.
\begin{figure}[htbp]
\begin{center}
\includegraphics[width=7.8cm]{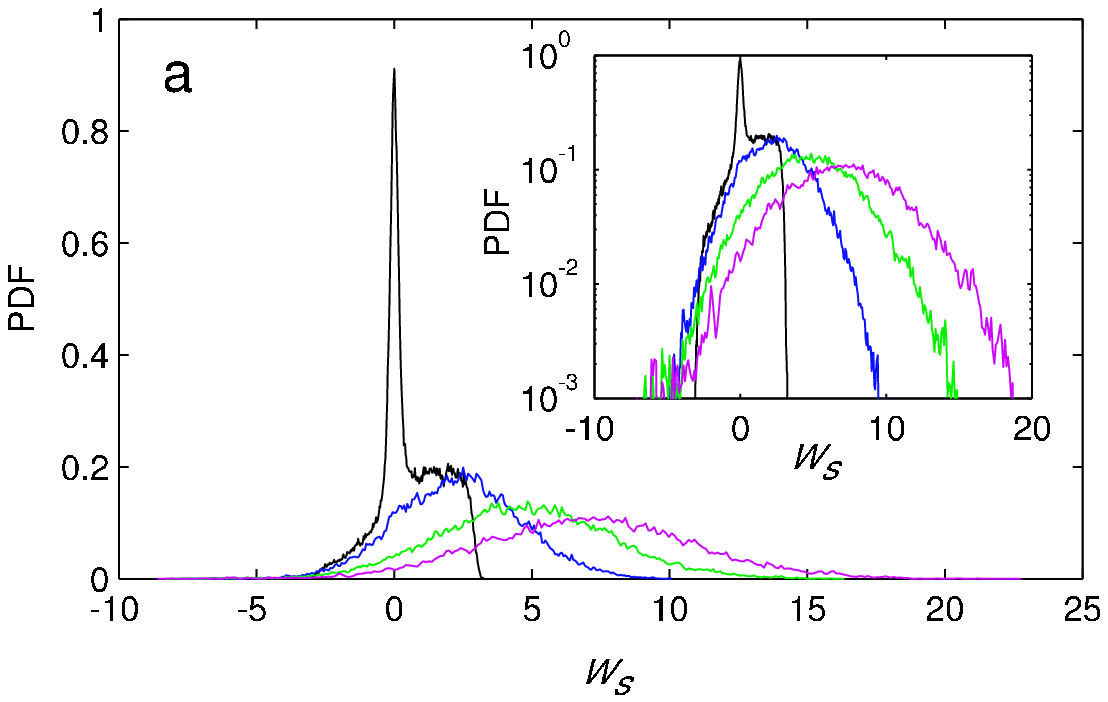}
\includegraphics[width=7.8cm]{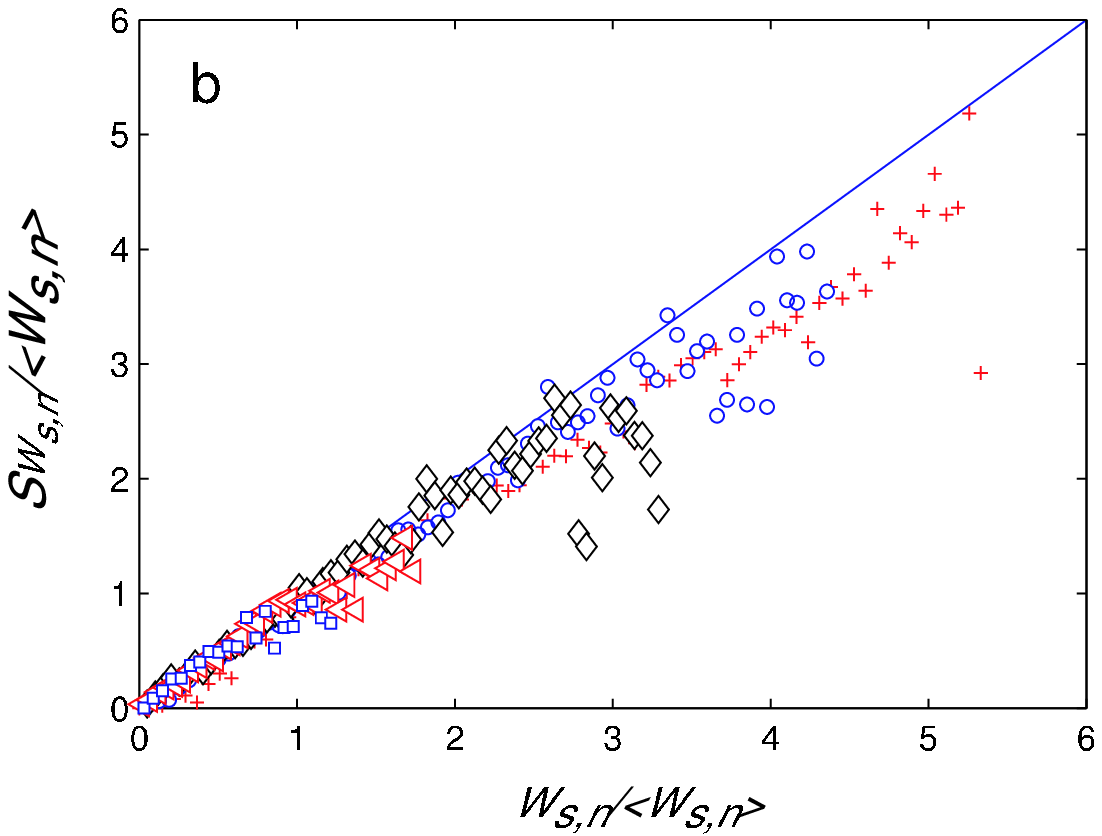}
\caption{a) Distribution of stochastic work for different number
of period $n=1$, $4$, $8$ and $12$ ($f=0.25$~Hz). Inset: Same data
in lin-log. b) Normalized symmetry function as function of the
normalized work for $n=1$ ($+$), $2$ ($\circ$), $4$ ($\diamond$),
$8$ ($\triangle$), $12$ ($\Box$).} \label{fig:wspdfn}
\end{center}
\end{figure}
We discuss in the following the PDFs $P(X_n)$ of $X_n$ where $X_n$
stands for one of the three variables  $W_{sn}$, $W_{cln}$ and
$Q_n$. For these quantities we also study the SSFT which states
that:
\begin{equation}
S_{X_n}=\log(\frac{P(X_n=w)}{P(X_n=-w)})\rightarrow {w}
\  {\rm for} \ n \rightarrow \infty \label{eq:SX} %\\
%\Sigma(n) \rightarrow 1 \  {\rm for} \ n \rightarrow \infty
\end{equation}
where $S_{X_n}$ is called the symmetry function. It is important
to notice that Eq.~\ref{eq:SX} for $W_{sn}$ should hold for all
$n$. Indeed taking into account that $Q_n=-\Delta U+W_{sn}$ it is
easy to realize that in this case $W_{sn}/T$ is just the total
entropy production defined by Seifert \cite{seifert07,seifert07b}
who has shown that that for this quantity the SSFT holds for any
integration time which is an integer number of periods of the
forcing.

We consider first $P(W_{sn})$ which is plotted in
Fig~\ref{fig:wspdfn}a) for various  $n$. The distribution presents
a sharp peak for $n=1$, which disappears as soon as $n$ increases.
The distribution then tends to a gaussian for high values of $n$.
To directly test the SSFT, we plot $S_{W_{sn}}/<W_{sn}>$ as a
funtion of $W_{sn}/<W_{sn}>$ in Fig.~\ref{fig:wspdfn}b).  It is
remarkable that straight lines are obtained even for $n$ close to
1, where the distribution presents a  very complex and unusual
shape. These curves seem to collapse on the curve $y=x$.
\begin{figure}[htbp]
\begin{center}
\includegraphics[width=6cm]{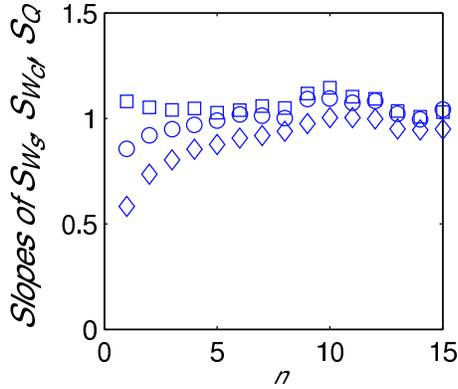}
\caption{Slope at the origin of the symmetry functions of $W_{sn}$
($\circ$) ,$W_{cln}$ ($\Box$) and $Q_n$ ($\diamond$) as function
of $n$ ($f=0.25$~Hz).} \label{fig:slopes}
\end{center}
\end{figure}
 We have looked closer to the slope of
$S_{W_{sn}}$ for small work fluctuations. Its evolution is shown on
the figure \ref{fig:slopes}.
 The SSFT is not verified at small $n$ but
 it is rapidly valid for increasing number of periods. The
 convergency turns out to be much faster than that observed in
 a harmonic oscillator  \cite{joubaud07} where SSFT is satisfied
 only for $n>30$. This is indeed an important point
because as we have already mentioned  $W_{sn}/T$ corresponds, in
this case,
 to the total entropy production,
 for which SSFT has to be verified for all $n$ \cite{seifert07b}.
Thus  the statistical and numerical uncertainties could explain
the fact that in our data and in those of ref.~\cite{saikia07} the
slope of $S_{W_{sn}}$ versus $W_{sn}$ for $n=1$ is not exactly
one. Indeed in ref.\cite{joubaud07b} it has been shown that even
in the gaussian case it is required a very large statistical
accuracy in the calculation of the total entropy to satisfy SSFT
for small $n$. In our case this accuracy has to be larger than in
the gaussian case because of the extremely complex shape of the
$P(W_{sn})$ for n=1.
\begin{figure}[htbp]
\begin{center}
\includegraphics[width=8.4cm]{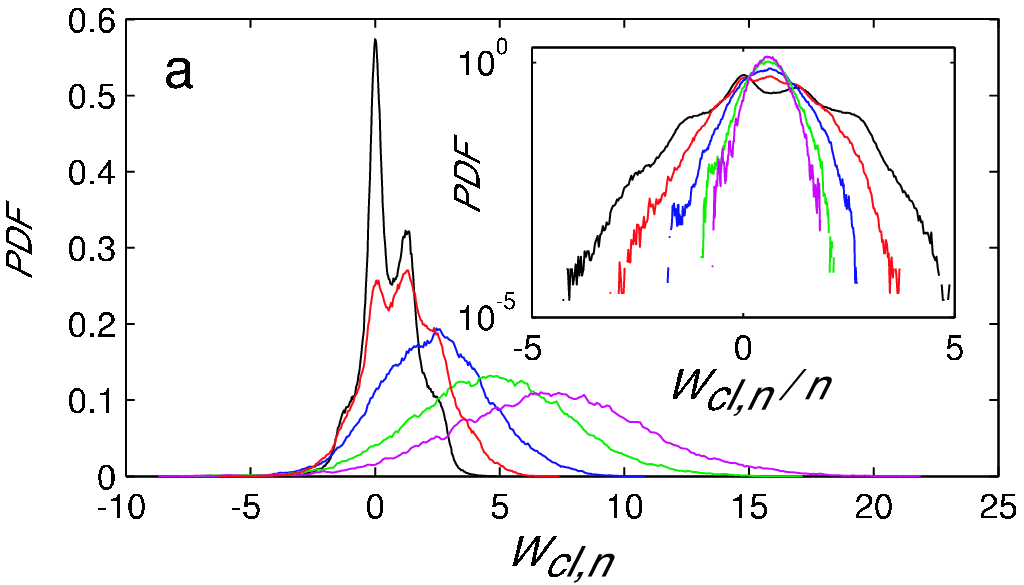}
\includegraphics[width=8.4cm]{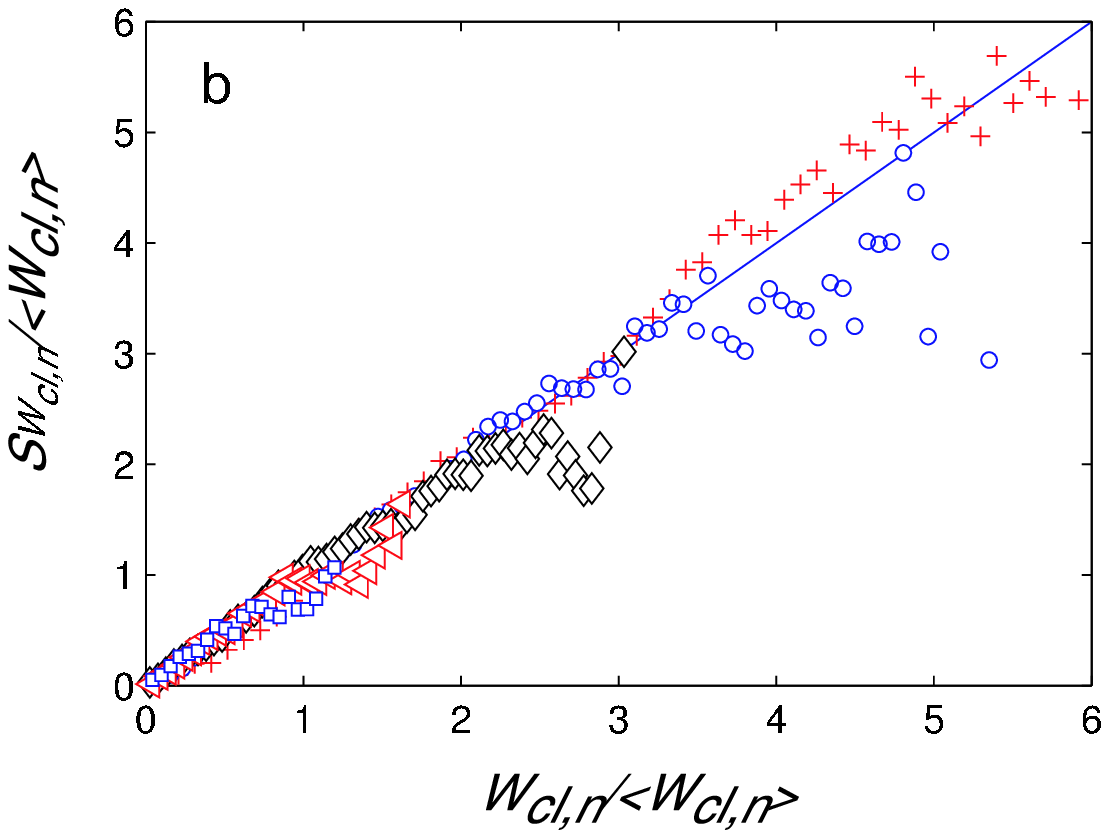}
\caption{a) Distribution of classical work $W_{cl}$ for different
numbers of period $n=1$, $2$, $4$, $8$ and $12$ ($f=0.25$~Hz).
Inset: Same data in lin-log. b) Symmetry function as function of
the normalized work (same symbols as in Fig. \ref{fig:wspdfn}.)}
%c) Slope at the origin of the symmetry functions of Ws ($\Box$), Wcl ($\circ$) and Q ($\diamond$) as function of $n$.}
\label{fig:wclpdfn}
\end{center}
\end{figure}
For the classical definition of the work $W_{cln}$, the
fluctuations are  larger than those  of $W_{sn}$
(Fig.~\ref{fig:wclpdfn}a).
 Again, the distributions tend to a gaussian for large
 $n$ (inset of Fig. \ref{fig:wclpdfn}a).
 On Fig.~\ref{fig:wclpdfn}b), we have plotted the normalized symmetry function of $W_{cln}$.
We can see that the curves
are close to the line of slope one. For high values of work, the
dispersion of the data increases due to the lack of events.
The slopes at the origin are shown in Fig.~\ref{fig:slopes}.
We can see an increase of the slope near zero and then an oscillation.
Although we can not average over more periods,
 the slope seems to tends toward 1 as expected by the SSFT. The
 strong analogy existing between  the convergency of $W_{sn}$ and
 $W_{cl}$ is probably due to the fact that, as already mentioned,
 they differ only for a boundary term that rapidly goes to zero
 when $n$ is increased.
\begin{figure}[htbp]
\begin{center}
\includegraphics[width=8.4cm]{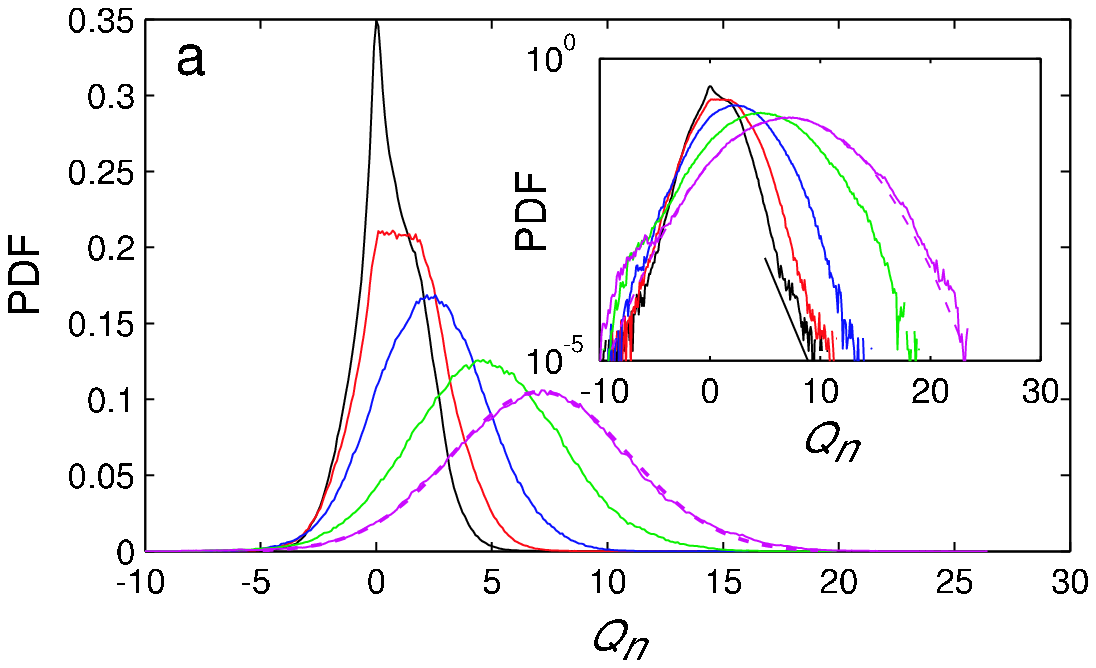}
\includegraphics[width=8.4cm,]{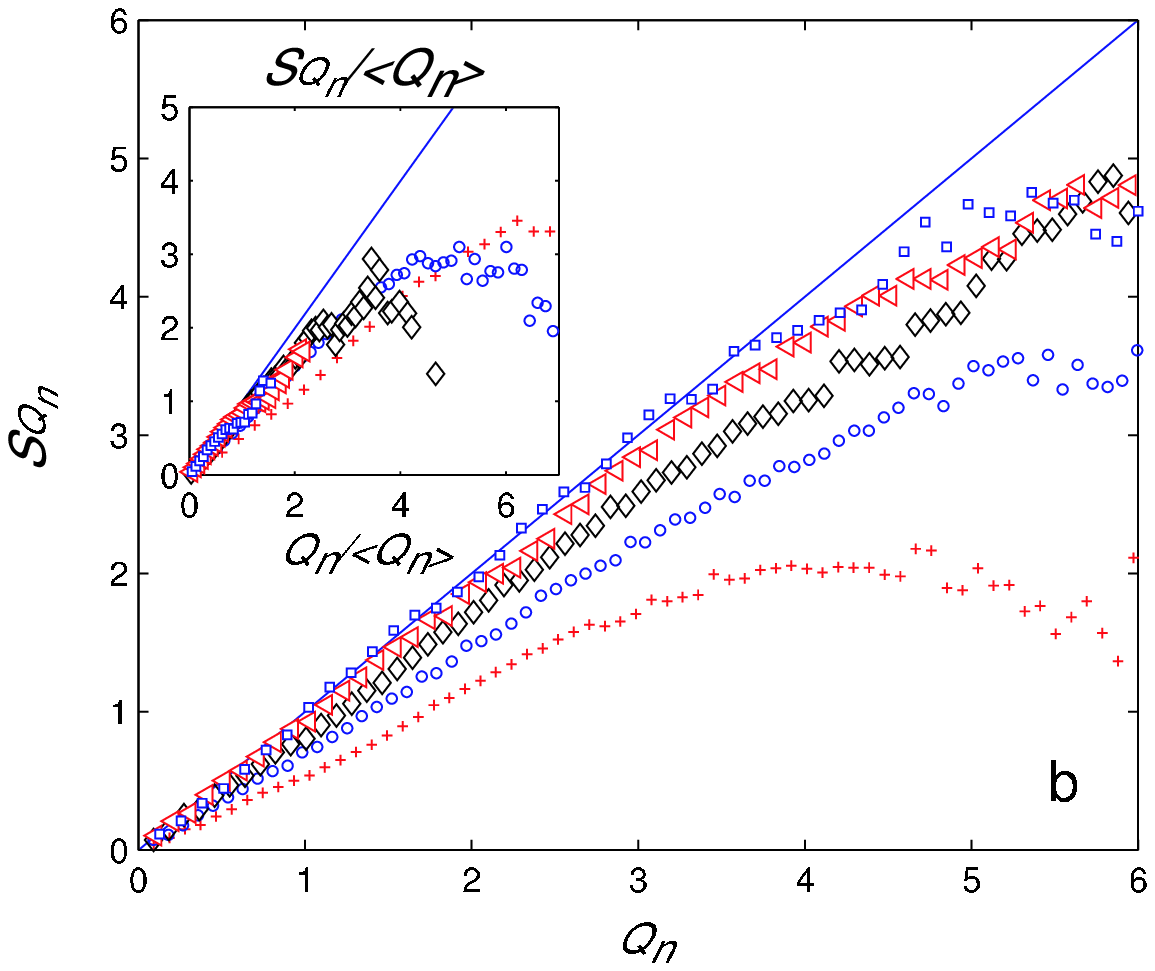}
\caption{a) Distribution of the dissipation for different numbers
of period $n=1$, $2$, $4$, $8$ and $12$. The straight line is
proportional to $exp(-Q_n)$. The dashed line is a gaussian fit of
the PDF for $n=12$.  Inset: Same data in lin-log . b) Symmetry
function of the dissipation for $n=1$, $2$, $4$, $8$ and $12$.
Inset: normalized symmetry function ($f=0.25$~Hz). (same symbols
as in Fig. \ref{fig:wspdfn}.)} \label{fig:disspdfn}
\end{center}
\end{figure}
Now, we consider the heat fluctuations.
The PDFs show even larger fluctuations (Fig~\ref{fig:disspdfn}).
We can notice first that the PDFs decrease exponentially in the asymptotic limit.
This is  due to the exponential
tails of the distribution of $\Delta U_0$ that become predominant.
This behavior is less pronounced for higher values
of $n$. Near the maximum, these distributions tend to be gaussian as shown by the exponential fit (dashed line).
On the figure \ref{fig:disspdfn}b), the symmetry function of the dissipated heat is shown as function
of the dissipated heat. We can define two regions, in the first one, the curve increases linearly and then bents.
On the inset of Fig.~\ref{fig:disspdfn}b), which shows the normalized symmetry function of $Q_n$, the two different regimes appear clearly.
First, the slope of the linear part increases with $n$ and goes asymptotically toward 1, as shown on Fig. \ref{fig:slopes}.
 Only the linear part for $Q_n<3$ has been fitted. Starting near 0.6 for one period, it grows toward 1 more slowly than for the classical work.
 This shows that the SSFT holds asymptotically when $n$ increases for $Q_n\ll\left<Q_n\right>$. On the other regime, one could expect the curve
to reach the value 2 from the shape of the exponential tails.
 However, their size are too
  small to create a horizontal asymptote.
   Still, when normalizing the data by the average
   of $Q_n$ (inset of Fig. \ref{fig:disspdfn}b),
   the data begin to explore the second regime around
   a value compatible with the value $2$,
   that would need longer measurements (over more period) to be confirmed.

In conclusion, we have experimentally investigated the power
injected in a bistable colloidal system by an external oscillating
force. We find that the injected power in the stochastic resonance
regime presents a maximum when the frequency of the driving force
corresponds to half of the Kramers' rate. We have compared the
stochastic work and the classical work for different frequencies
and shown that although the average values of each work are equal,
the distributions of work reveal some differences. They present
large tail toward negative values and their shape are non-gaussian
close to the resonance when averaging over a single period.
 Analyzing in more details the distribution of work and dissipated heat,
 we have tested the validity of SSFT for a non-linear potential.
 We have shown that FT rapidly converge to the asymptotic value for rather small $n$.
The fact that for the total entropy $W_{sn}/T$, SSFT is not
satisfied exactly for small $n$ is certainly due to statistical
and numerical inaccuracy. We have also shown that  the SSFT is
only valid for small values
  of dissipated heat compared to the mean value at long time.
  It would be interesting to change the symmetry
  of the driving force cycle \cite{mai07} to explore the limit of the
  FT.

We acknowledge useful discussion with R. Benzi, K. Gawedzki and S.
Joubaud. This work has been partially supported by
ANR-05-BLAN-0105-01.

\end{document}